\documentclass[twoside,prb,letterpaper,preprintnumbers,superscriptaddress,notitlepage,onecolumn]{revtex4}

\usepackage{mathptmx}
\usepackage{helvet}
\usepackage{graphicx,fancyhdr}% Include figure files
\usepackage{dashbox,color}%
\usepackage{bm}% bold math
\usepackage{sidecap}%
\definecolor{lightblue}{rgb}{0.6,0.9,1}
\definecolor{myrefblue}{rgb}{0.1,0.6,1}
\definecolor{myblue}{rgb}{0,0,0}
\definecolor{nmat}{rgb}{0.7,0.04,0.26}

\begin{document}

\pagestyle{fancy}

\renewcommand{\headrule}{\vskip-3pt\hrule width\headwidth height\headrulewidth \vskip-\headrulewidth}

\fancypagestyle{plainfancy}{%
\lhead{}
\rhead{}
\chead{}
\lfoot{}
\cfoot{}
\rfoot{\bf\scriptsize\textsf{\thepage}}
\renewcommand{\headrulewidth}{0pt}
\renewcommand{\footrulewidth}{0pt}
}

\fancyhead[LE,RO]{}
\fancyhead[LO,RE]{}
\fancyhead[C]{}
\fancyfoot[LO,RE]{}
\fancyfoot[C]{}
\fancyfoot[LE,RO]{\bf\scriptsize\textsf{\thepage}}

\renewcommand\bibsection{\section*{\sffamily\bfseries\normalsize {References}\vspace{-10pt}\hfill~}}
\newcommand{\mysection}[1]{\section*{\sffamily\bfseries\normalsize {#1}\vspace{-10pt}\hfill~}}
\renewcommand{\subsection}[1]{\noindent{\bfseries\normalsize #1}}
\renewcommand{\bibfont}{\fontfamily{ptm}\footnotesize\selectfont}
\renewcommand{\figurename}{Figure}
\renewcommand{\refname}{References}
\renewcommand{\bibnumfmt}[1]{#1.}

\newcommand{\be}{\begin{equation}}
\newcommand{\ee}{\end{equation}}
\newcommand{\bn}{\begin{eqnarray}}
\newcommand{\en}{\end{eqnarray}}
\newcommand{\ii}{\'{\i}}
\newcommand{\ca}{\c c\~a}

\makeatletter
\long\def\@makecaption#1#2{%
  \par
  \vskip\abovecaptionskip
  \begingroup
   \small\rmfamily
   \sbox\@tempboxa{%
    \let\\\heading@cr
    \textbf{#1\hskip1pt$|$\hskip1pt} #2%
   }%
   \@ifdim{\wd\@tempboxa >\hsize}{%
    \begingroup
     \samepage
     \flushing
     \let\footnote\@footnotemark@gobble
     \textbf{#1\hskip1pt$|$\hskip1pt} #2\par
    \endgroup
   }{%
     \global \@minipagefalse
     \hb@xt@\hsize{\hfil\unhbox\@tempboxa\hfil}%
   }%
  \endgroup
  \vskip\belowcaptionskip
}%
\makeatother

\thispagestyle{plainfancy}

\fontfamily{helvet}\fontseries{bf}\selectfont
\mathversion{bold}
\begin{widetext}
\begin{figure}
\vskip0pt\noindent\hskip-0pt
\hrule width\headwidth height\headrulewidth \vskip-\headrulewidth
\hbox{}\vspace{4pt}
\hbox{}\noindent\vskip10pt\hbox{\noindent\huge\sffamily\textbf{Microscopic 
Description of Unconventional Nodal}}\vskip0.05in\hbox{\noindent\huge\sffamily\textbf{Superconductivity in FeSe}}

\vskip10pt
\hbox{}\noindent\begin{minipage}{\textwidth}\flushleft
\renewcommand{\baselinestretch}{1.2}
\noindent\hskip-10pt\large\sffamily Mukul S. Laad,$^{1,2}$ 
Byron Freelon,$^{3,4}$ \& Luis Craco$^{5,6}$

\end{minipage}
\end{figure}
\end{widetext}

\begin{figure}[!h]
\begin{flushleft}
{\footnotesize\sffamily
$^1$Institute of Mathematical Sciences, Taramani, Chennai 600113 and Homi 
Bhabha National Institute, Anushakti Nagar, Trombay, Mumbai 400085, India. 
$^2$ Helmholtz-Zentrum Berlin f\"ur Materialien und Energie, 
Lise-Meitner-Campus, A-307,Hahn-Meitner-Platz 1, 14109 Berlin, Germany.
$^3$ Department of Physics, University of Louisville, Louisville, KY 40208, 
USA. $^4$Department of Physics, Massachusetts Institute of Technology, 
Cambridge, Massachusetts, 02139, USA. $^5$Instituto de F\ii sica, Universidade 
Federal de Mato Grosso, 78060-900, Cuiab\'a, MT, Brazil. $^6$IFW Dresden, 
Institute for Solid State Research, P.O. Box 270116, D-01171 Dresden, Germany}
\end{flushleft}
\end{figure}

%\small
\fontsize{9pt}{8pt}\selectfont
\renewcommand{\baselinestretch}{0.9}
%abstract
\noindent\sffamily
{Finding of unconventional superconductivity (USC) in FeSe in an electronic 
``normal'' state with broken $C_{4v}$ symmetry testifies to the diversity 
of pairing states in Fe-based superconductors. Moreover, such USC emerges 
as a direct instability of a normal state without Landau Fermi liquid 
quasiparticles, increasingly dubbed a ``strange''metal.~\cite{sachdev}  
Here, we combine inputs from a first-principles correlated electronic 
structure method (LDA+DMFT) and symmetry analyses to propose a novel 
mechanism for unconventional nodal superconductivity as a direct 
instability of an incoherent bad metal without Landau Fermi-liquid 
quasiparticles.  We find that a ferro-quadrupolar order, with novel spin 
quadrupolar correlations enhances orbital-selective Mottness in FeSe, 
and competes with unconventional, nodal superconductivity with $s_{\pm}$ 
pair symmetry.  We support our proposal by demonstrating good accord with 
spectral and magnetic fluctuation data, and rationalize the strain and 
pressure dependence of $T_{c}$ by appealing to competition between 
superconductivity and electronic ferro-quadrupolar order.}

\mathversion{normal}
\normalfont\normalsize

\section{Introduction}
Unconventional Superconductivity (USC) in Iron-based pnictides and 
chalcogenides can seemingly host a range of gap function 
symmetries.~\cite{johnson} a surprising fact that seems to be related to 
their intrinsically multi-orbital (MO) character (in contrast to cuprates, 
where $d$-wave SC is universal). The essential idea is that suitable external 
perturbations sensitively modify the anisotropic band structure in a 
band-selective way, and it is then conceivable that SC gap nodes may or 
may not exist, depending ultimately upon whether or not the SC gap function 
$\Delta({\bf k})$ intersects the multi-sheeted, {\it renormalized} Fermi 
surface. Finally, as in cuprates and some $f$-electron systems, USC is now 
found to generically arise from a non-Landau Fermi liquid 
(nLFL)~\cite{sachdev} metal near optimal doping, precluding description of 
USC as a conventional instability of an itinerant LFL metal.

The ``simplest'' FeSe system is an especially suitable case in point. 
Additional unique features here are:  $(i)$ SC (reaching $T_{c}=37$~K 
under high pressure) and with a record high $T_{c}\simeq 100$~K in FeSe 
films deposited on SrTiO$_{3}$ substrates~\cite{FeSe-record-Tc}
 occurs in a local moment metal~\cite{cava} without 
antiferromagnetic (AF) order (though AF order appears at high pressure), 
in contrast to other cases where SC at ambient pressure always occurs 
close to AF order.  
$(ii)$ USC occurs in a ``normal'' state 
which is simultaneously an incoherent nLFL metal~\cite{laad,kotliar} and 
breaks the fourfold discrete lattice rotational symmetry,~\cite{song} as 
in some underdoped 122-FeAs systems.~\cite{fisher} Whether $(i)$ this 
``normal'' state thus hosts orbital nematic~\cite{phillips} order, and 
its relation (competitor or facilitator) to SC, and $(ii)$ do they 
originate from the same set of effective interactions in the ``normal'' 
incoherent metal, are intriguing issues.  $(iii)$ Tunneling data reveal
nodal-SC for pure FeSe which, remarkably, evolves into nodeless-SC with
Te substitution ($x$).~\cite{song} $(iv)$ Tensile strain~\cite{brahimi} 
and minute Fe-excess~\cite{mcqueen} rapidly suppress SC, which, at first 
glance, is not inconsistent with multiple orbitals and gap nodes.     

Taken together, $(i)-(iv)$ imply challenging constraints for a theory of 
SC in FeSe$_{1-x}$Te$_{x}$: $(i)$ poses an obvious challenge to theories 
relying on AF quantum criticality, since AF is only found in Te-rich 
and/or pressurized 
samples, though this does not exclude strong, short-range dynamic spin 
fluctuations.~\cite{cava} As pointed out above, both, the nature and 
extent of orbital-dependent electronic structure reconstruction (O-ESR) 
in FeSe is crucial to distinguish nodeless versus nodal SC. That the 
observed anisotropy is much greater than its pure structural 
counterpart~\cite{song} favors a predominantly electronic origin 
for O-ESR.  In light of $(i)-(ii)$, understanding $(iii)$ must now 
involve interplay between the form-factor of $\Delta({\bf k})$ and changes 
in the anisotropic {\it correlated} electronic structure with $x$.

In this work, we develop a theory to address these issues. In a 
non-trivial extension of earlier work,~\cite{laad2} we study the origin 
and consequences of the two-fold ($C_{2v}$) symmetry in the normal and 
SC states~\cite{hofmann} in FeSe. In fact, for FeSe, one has to look 
for USC in the $C_{2v}$ structure, a difference that has important 
implications for pair symmetry. Earlier (Hartree-Fock) Bogoliubov 
approach~\cite{wu} includes $C_{4v}\rightarrow C_{2v}$ symmetry breaking 
phenomenologically. Our microscopic approach involves deriving  both 
orbital nematic (ON) and USC orders from a specific residual interaction 
that we argue becomes more relevant than the incoherent one-particle 
mixing in a nLFL metal.~\cite{pwa,laad}  

Specifically, two LDA+DMFT studies~\cite{craco-fese,kotliar} find a 
qualitatively similar ``normal'' state: an incoherent nLFL metal without 
LFL quasiparticles, argued to occur either via a lattice orthogonality 
catastrophe~\cite{craco-fese} via spin-orbital freezing in a metallic, 
orbital-selective Mott phase (OSMP) or via the related Hund's 
metal~\cite{kotliar1} route.  Already in the ``normal'' state at high $T$, 
it is known~\cite{kotliar} that orbital-selective Mott correlations 
selectively localize the $xy$ states.  What is important for our purposes 
here is that the resulting orbital-selective (OS) Mott state and reconstruction 
of LDA Fermi surfaces (FS) breaks the analytic continuation with any 
renormalized FL metal that results from the unreconstructed LDA Fermi 
surfaces (which is indeed what would happen if local multi-orbital 
correlations were weak and selective-Mott localization of $xy$-staes would {\it not} occur). We argue that the resultant irrelevance of 
coherent one-electron mixing ($H_{hyb}^{(1)}$) the OSMP that occurs at 
intermediate-to-strong coupling precludes the ``weak coupling'' BCS-like 
instability to a superconductor.  In earlier work,~\cite{laad2} we argued 
that these now arise via a route first expounded by Anderson~\cite{pwa} in 
the cuprate context: the essential idea is that upon coupling two Luttinger 
liquids (LLs), coherent one-electron hybridization scales to irrelevance 
because of spin-charge separation, but, to higher (second) order, spinon 
or holon pairs can coherently tunnel between two LLs.  Since the ``normal'' 
state we find is a {\it local} non-Landau FL metal, it is the residual 
{\it inter-site} interaction (formally obtained by repacing LL chains in Anderson's proposal with local {\it sites} in our case) which is preferentially 
relevant relative 
to the one-electron interband hybridization, and generates the instabilities 
to ordered states directly from the incoherent metal (see below).  In other 
words, in analogy with Anderson's idea, irrelevance of $H_{hyb}^{(1)}$ itself 
makes the corresponding two-particle hopping (appearing as a residual 
interaction, $H_{res}$) more relevant. Interestingly, the same $H_{res}$ also 
generates an (electronic) ON instability in the competing particle-hole 
(p-h) channel, allowing us to study their interplay. 

\section{Theory and results}

The Fermi pockets of tetragonal-FeSe in the local-density approximation (LDA) 
with $C_{4v}$ symmetry are predominantly composed of the Fe-$d_{xz,yz,xy}$ 
orbital states: the central hole pocket at $\Gamma=(0,0)$ point is mainly 
of ${yz,xz}$ character, while the electron pockets at 
$M=(\pm\pi,0),(0,\pm\pi)$ have mainly $xy,xz$ orbital component. 
For simplicity, we focus on the $xz,yz$ orbitals: due to interband proximity 
effect,~\cite{laad2,wu} the remainder of the $d$ orbitals will also play 
a role in reality.  We first show how an electronic (orbital) nematic drives 
FeSe into an OSMP with co-existent localized ($xz$) and bad-metallic ($yz$) 
states.  The one-electron hopping matrix in this orbital sector, formally 
written as $H_{hyb}=\sum_{i,j,a,b,\sigma}t_{ab}(c_{ia\sigma}^{\dag}c_{jb\sigma}+H.c)$, 
with $i,j$ being nearest (n.n) and next-near neighbors n.n.n), is strongly 
geometrically frustrated (GF) in the FeAs(Se) systems with a large ratio 
of n.n.n (diagonal) to n.n hopping strengths: 
$t_{ab}^{(nnn)}/t_{ab}^{(nn)}\simeq 0.7-1.0$.  As discussed before,~\cite{laad2} 
residual two-particle interactions, i.e, hopping processes to second-order 
in $H_{hyb}^{(1)}$, are more relevant in the non-LFL ``normal'' state.  
As derived before,~\cite{laad2} this reads

\be
H_{res} \simeq  -\frac{1}{U'+J_{H}}\sum_{<i,j>,a,b,\sigma,\sigma'} 
t_{ij,a,b}^{2} (c_{i,a,\sigma}^{\dag}c_{j,b,\sigma} 
c_{j,b,\sigma'}^{\dag}c_{ia\sigma'} \nonumber + H.c)\;.
\ee
This resembles a generalized {\it effective} spin-orbital superexchange, now 
ipso-facto validated by the selective-Mott nature of the ``normal'' state.
It is important to emphasize that this reflects the inherent spin-orbital 
entanglement in multi-orbital correlated systems, and implies that the 
spin-fluctuations are intrinsically orbital-selective: in particular, onset 
of an OSMP will enhance anisotropy of spin fluctuations.  Thus, the form of 
$H_{res}$ shows that entangled short-ranged dynamical spin-orbital correlations 
facilitate instabilities of the incoherent ``normal'' state to competing 
orders.  This is because   $H_{res}$ can now be readily decoupled into $(a)$ 
particle-hole (p-h) and $(b)$ particle-particle (p-p) mean fields, a procedure 
exact within dynamical mean-field theory (DMFT), since both scale as $1/D$. 
Here, $(a)$ corresponds to either ON ($\sigma=\sigma'$) or AF-SDW 
($\sigma'=-\sigma$) orders, while $(b)$ represents an USC instability. 
Since SC in FeSe arises {\it after} the lattice distortion (implying ON) 
has set in, we adopt the stratagem of first studying the ON instability, 
and then focus on the USC instability, with very interesting consequences. 
In momentum space, $H_{res}$ looks very suggestive:

\be
\nonumber
H_{res} \simeq  -\sum_{k,k',a,b,\sigma,\sigma'}V_{ab}\gamma(k)\gamma(k')c_{k,a,\sigma}^{\dag}(\delta_{kk'}\delta_{\sigma\sigma'} - 
c_{-k,b,-\sigma}^{\dag}c_{-k',b,-\sigma'})c_{k',a,\sigma'} \;,
\ee
where $\gamma(k)=(c_{x}+c_{y})+\alpha c_{x}c_{y}$~\cite{mazin,laad,wu} and 
$c_{a}=$cos$k_{a}$, with $a=x,y$.  $\alpha=\simeq O(0.7)$ is large, 
thanks to the strong GF in FeSe, and $V_{ab}=\frac{t_{ab}^{2}}{U'+J_{H}}$. 
This can be decoupled as

\be
\nonumber
H_{res}^{(MF)} = -\sum_{k,k',a,b,\sigma,\sigma'}V_{ab}\gamma(k)\gamma(k')[(1-\langle n_{k'b\sigma'})\rangle n_{k,a,\sigma}  + \langle n_{ka\sigma}\rangle n_{k'b\sigma'} 
- \langle c_{k,a,\sigma}^{\dag}c_{-k,b,-\sigma}^{\dag}\rangle c_{-k,b,-\sigma}c_{k,a,\sigma} 
+ H.c] \;.
\ee
ON order arising as a consequence of ferro-orbital order (FOO) directly 
arises from the first term in $H_{res}^{(MF)}$ above, which pushes the 
$b$-fermion ($xz$-orbital) states below the $a$-fermion ($yz$-orbital) 
states by an amount $-V_{ab}\sum_{k}\gamma(k')\langle n_{k'b\sigma'}\rangle$. The 
resulting FOO is described by an order parameter, $Q_{i,z}=(n_{i,xz}-n_{i,yz})$, 
which is precisely the ferro-quadrupolar order (FQO) parameter.~\cite{blumberg} 
$Q_{i,z}$ couples to an appropriate symmetry-adapted phonon mode and drives 
$C_{4v}\rightarrow C_{2v}$ symmetry breaking and to orthorhombicity: thus, the 
$T$-dependence of the FQO is reflected in that of the orthorhombicity, 
$<O>=\frac{b-a}{b+a}$.  Furthermore, the fact that it shows a mean-field 
$T$-dependence is also in accord with our view, since we derive this 
instability by a mean-field decoupling of $H_{res}$. USC arises from the 
third term. The second term in $H_{res}^{MF}$ is especially interesting: it 
reads $H_{res,ph}^{(MF)}=\sum_{a=xz,yz}\Delta_{ph}^{a}\sum_{k}\gamma(k)n_{a}(k)$ with 
$\Delta_{ph}^{a}=\sum_{k'}\gamma(k')\langle n_{a}(k')\rangle$, which is 
precisely the bond-orbital order (BOO) needed to reconcile ARPES 
dispersions in the electronic nematic phase in FeSe.~\cite{jiang,ding} Thus, 
remarkably, this specific feature of the ON state in FeSe (and possibly in 
other Fe pnictides as well) falls out naturally as a consequence of the 
instabilities arising from residual two-particle (inter-site and 
intra- as well as inter-orbital) interactions in the incoherent nLFL 
``normal'' state we find.  We defer detailed comparison with FS to 
future work, but emphasize here that our proposed mechanism is naturally 
consistent with very recent ARPES work.~\cite{rhodes} A related proposal 
has recently been made within a slave-rotor approach,~\cite{si1} who use 
the same terms as those in our $H_{res}^{(MF)}$ phenomenologically.  Here, 
this term (precisely the last term in Eq.~(4) of Yu {\it et al.}~\cite{si1}) 
is derived from a leading-order two-particle residual interaction 
generated from an incoherent ``normal'' state.

\begin{figure}[!b]
\vskip50pt
\end{figure}
\begin{figure*}[!t]
\includegraphics[width=5.9in]{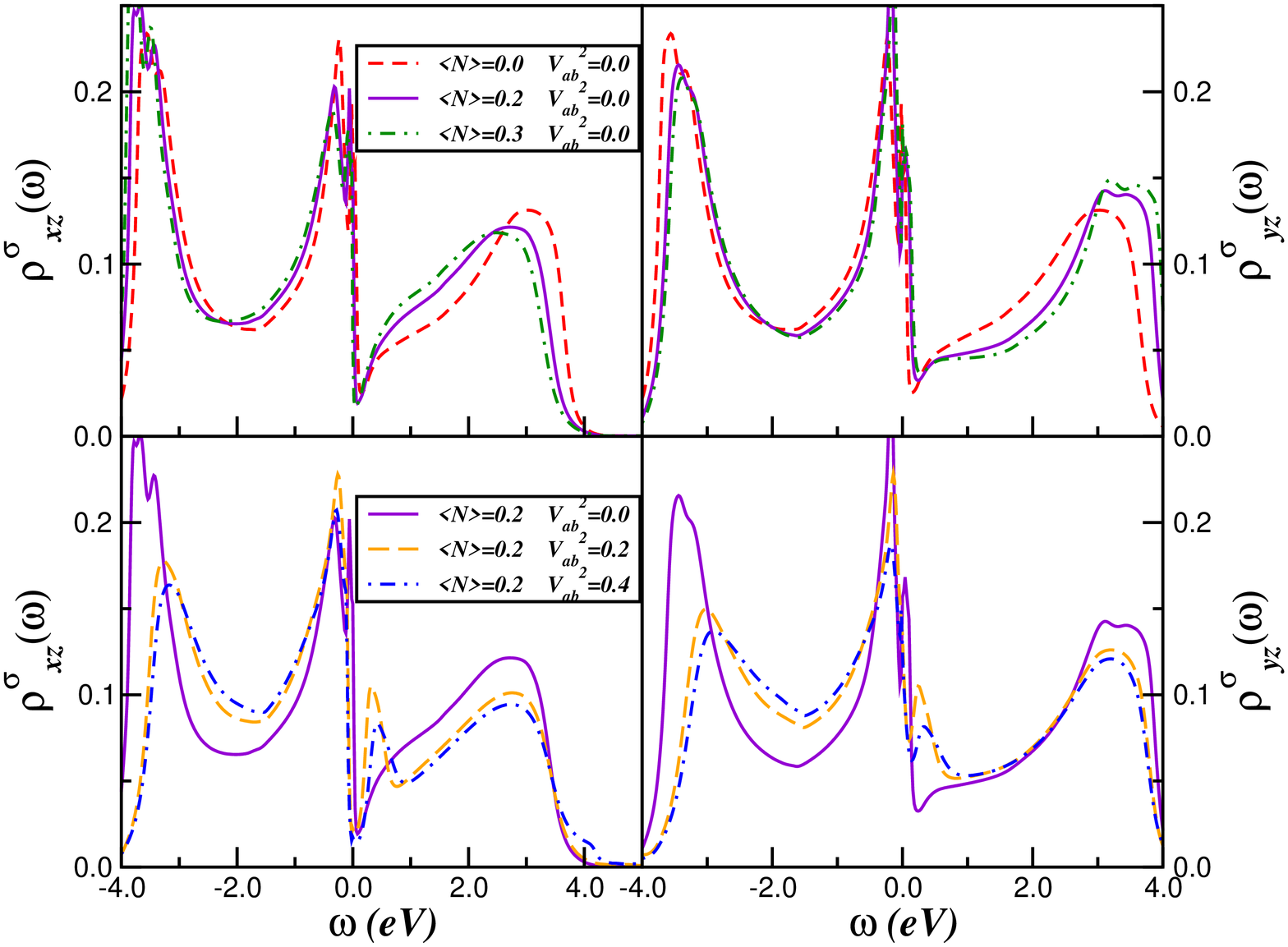}
\caption{
LDA+DMFT many-body density-of-states for $xz,yz$ bands in the orbital 
nematic (ON) (upper panels) and the orbital-nematic-plus-anisotropic 
extended $s_{\pm}$-wave superconducting (ON+SC) state proposed in the 
text.  Clear difference in $\rho_{xz}(\omega),\rho_{yz}(\omega)$, 
implying breaking of $C_{4v}$ symmetry down to $C_{2v}$ by ferro-orbital 
order (FOO) is visible. Further, the Bogoliubov quasiparticle DOS (lower 
panels) shows clear orbital-dependent $V$-shaped and gapless structure, 
consistent with in-plane gap nodes in FeSe.}
\label{fig1}
\end{figure*}

That this on-site FOO plus BOO, or a FQO state is the main competitor to USC, 
at least in FeSe, is also bared by generalizing the argument of Podolski 
{\it et al.}:~\cite{kee} The pseudospin generator 
$\Delta^{-}_{s-sc}=\frac{1}{\sqrt{2}}\sum_{k,a=xz,yz}c_{ka\uparrow}c_{-k,a,\downarrow}$ 
rotates the bond-order parameter $\Delta_{ph}^{(a)}$ above to the $s_{\pm}$-SC 
order parameter as $[\Delta^{-}_{s-sc}, \Delta_{ph}^{(a)}]=\Delta^{-}_{s_{\pm}}$: 
these orders thus compete (a reflection of both arising from the same  
$H_{res}$).  Within the $xz-yz$ orbital sector, the above relation also 
suggests the emergence of a hidden SO(6) invariance at a transition 
between ON+BOO and USC states. Remarkably, terms which break this emergent 
symmetry, such as finite chemical potential (doping) and strain (which 
produces explicit finite orbital polarization) can lead to a transition 
between ON+BOO and USC orders: these may already have been 
seen.~\cite{brahimi} In FeSe, ON order pushes the $xz$ ($b$) band below 
its ``normal'' state value by $-V_{ab}$ (first term in $H_{res}$ above).  
The non-existence of the $d_{xz}$ Fermi pocket in ARPES~\cite{ding} is a 
Fermi surface {reconstruction,~\cite{laura} arising from a p-h 
order-induced downshift of the $xz$-band states as above. This also 
naturally accounts for a Lifshitz transition at $T^{*}$.  Indeed, many 
observations~\cite{grinenko} have been interpreted in terms of a Lifshitz 
transition: we posit that this is associated with enhancement of orbital 
selectivity by orbital nematicity~\cite{suchitra,si1} arising from FOO+BOO 
as above. Very recent ARPES data~\cite{rhodes} are completely consistent 
with this result as well: the $xz$-orbital states are {\it not} seen in 
data and, furthermore, using only the $yz$-Fermi pocket to compute ARPES 
spectra in the USC state agree well with data as well. In fact, the USC 
scales with the $d_{yz}$-orbital spectral weight. Again, while this is 
put in by postulating an orbital polarization, such a term, driving a 
Lifshitz transition via enhanced orbital selectivity, arises from residual 
interactions ($H_{res}$) in our work, and involves mutually entangled, 
intersite and interband spin-orbital correlations (not spin-orbit coupling). 
These features explicitly break SO(6) symmetry,~\cite{kee} stabilizing 
ON+BOO order (without such symmetry breaking USC would always win over 
ON+BOO): thus, the onsite FOO, now arising from residual interactions, 
creates the proper conditions for emergence of BOO that competes with 
USC. This finding may have much broader applicability: in cuprates, 
recent work also indicates a bond-modulated order that may be the leading 
competitor to $d$-wave SC, at least in hole-doped cuprates.~\cite{subir2}

\begin{figure}[!b]
\vskip50pt
\end{figure}
\begin{figure*}[!t]
\includegraphics[width=5.9in]{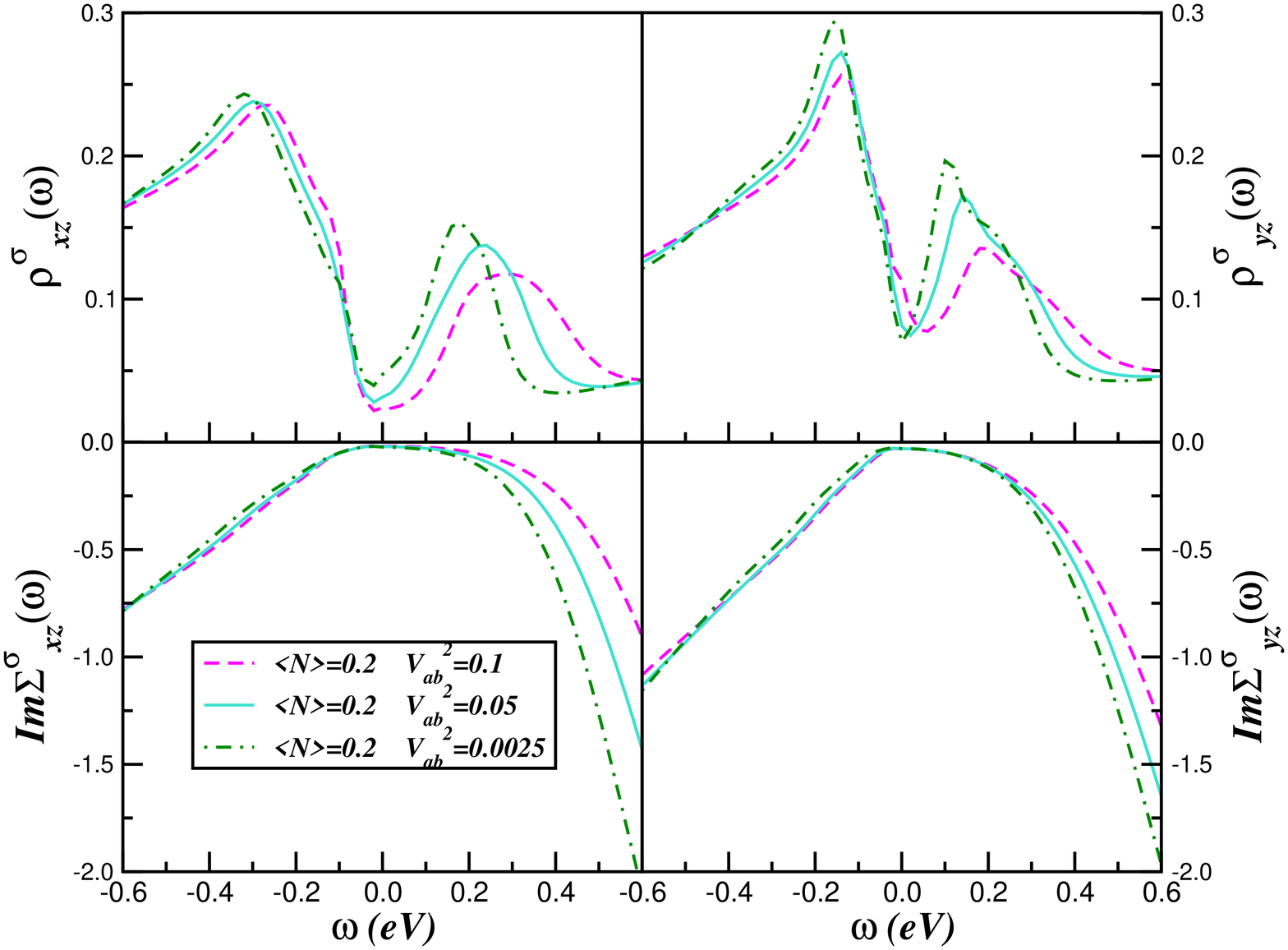}
\caption{
Orbital-resolved spectral functions (upper panels) and self-energies 
(lower panels). Direct comparison with normal state results~\cite{craco-fese} 
clearly shows how USC gap opening suppresses normal state incoherence, leading 
to a much more ``coherent'' self-energy with negligible damping at low energy. 
Thus, this shows that the ``normal''-USC phase transition is a 
coherence-restoring one, in analogy with cuprates.}
\label{fig2}
\end{figure*}

However, since USC now occurs in the ON phase with $C_{2v}$ symmetry, it 
is clear that the SC pair-symmetry cannot anymore be cleanly labeled by the 
$A_{1g}$ irreducible representation (irrep) of $C_{4v}$.  Instead, since the 
$A_{1g}$ irrep of $C_{2v}$ transforms like $x^{2},y^{2},z^{2}$,~\cite{hamermesh} 
it now follows that the $k$-space form-factor of the symmetry-allowed SC 
pair-function must rigorously be given by $\Delta_{\pm}({\bf k})=\Delta_{1}($cos$k_{x}+\eta$cos$k_{y})+\Delta_{2}$cos$k_{x}$cos$k_{y}$ with $\eta\neq 1$.  
Remarkably, this is precisely the ``anisotropic'' $s_{\pm}$ form assumed 
phenomenologically~\cite{wu}: our central result is that this is derived 
from a specific residual interaction in the incoherent metal that is also 
consistent with the spontaneous lowering of crystal symmetry in the 
``normal'' state. Interestingly, for $\alpha\simeq 0.7-1.0$, the anisotropic 
SC pair function will intersect the electron Fermi pockets~\cite{wu,chainani} 
at the $M$-points, accounting for signatures of nodal pairing in FeSe. 
Finally, examination of LDA FS shows small $c$-axis warping, so that 
accidental $c$-axis gap nodes cannot exist in FeSe. Thus, planar nodal 
$s_{\pm}$ SC emerges as the most probable in FeSe.  

We now show how our results  are in good semiquantitative accord with a host 
of spectral and magnetic data. The first term in $H_{res}$ directly leads to 
$\langle n_{xz}\rangle >\langle n_{yz}\rangle$ by itself, implying a FOO along 
with a finite orbital nematicity, 
$\langle N\rangle =\frac{n_{xz}-n_{yz}}{2(n_{xz}+n_{yz})}>0$
as a leading instability of the nLFL metal.  $\langle N\rangle >0$ 
immediately couples to an appropriate symmetry-adapted phonon mode, 
lowering the $C_{4v}$ crystal symmetry to $C_{2v}$.  This is explicitly 
manifested in the orbital-resolved DOS in Fig.~\ref{fig1}, which shows 
an enhanced pseudogap in $\rho_{xz}(\omega)$ at the expense of a reduced 
one in $\rho_{yz} (\omega)$; i.e, orbital-selective incoherence is enhanced 
in the ON phase.  Though this is already consistent with STM,~\cite{loidl} 
more studies are certainly needed to put this on surer ground.

\begin{figure}[!b]
\vskip50pt
\end{figure}
\begin{figure*}[!t]
\includegraphics[width=5.9in]{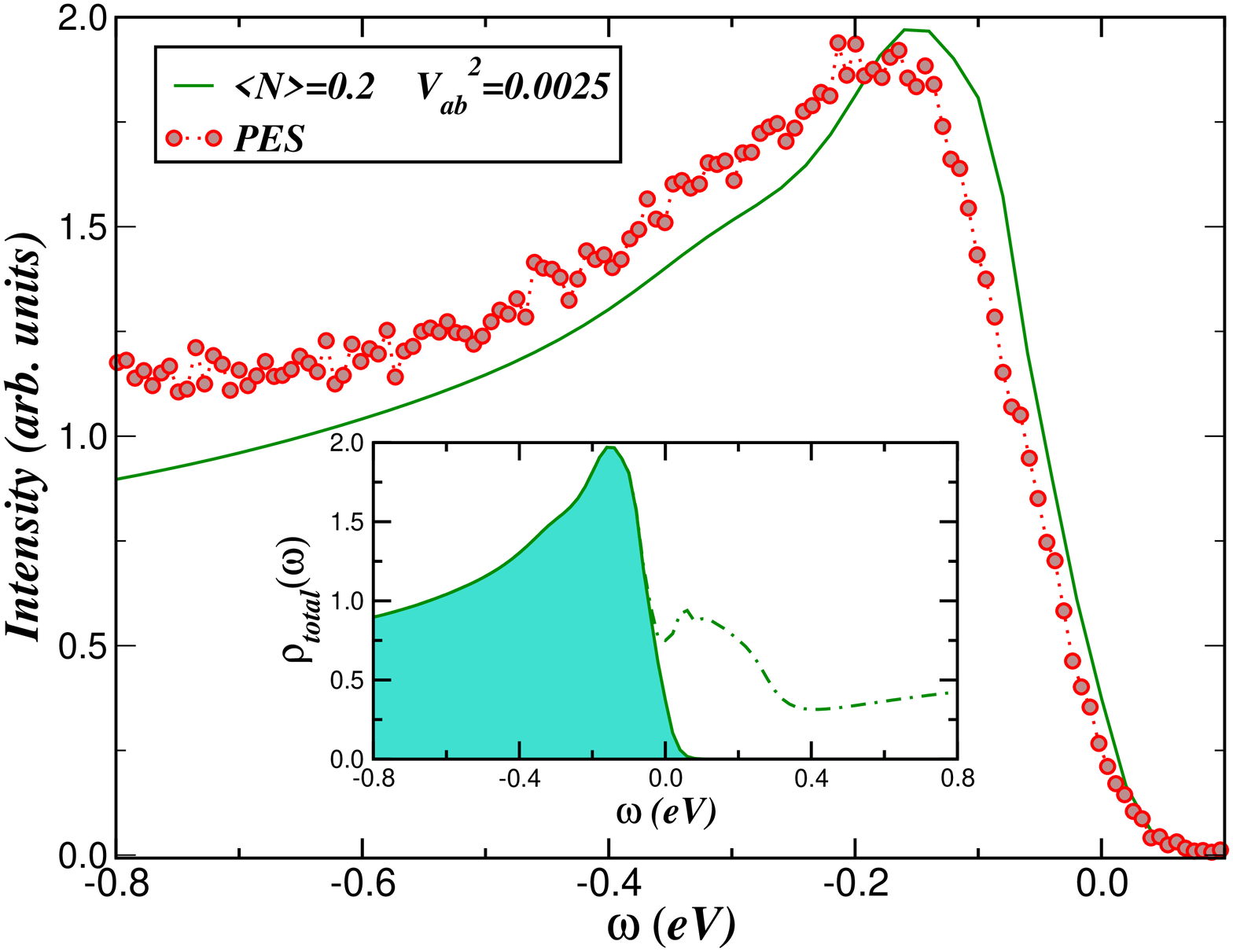}
\caption{
Comparison between the experimental angle-integrated photoemission 
(AIPES)~\cite{SC-FeSe-pes} and the total Bogoliubov quasiparticle DOS 
in the SC state. Very good quantitative accord between the two lends 
 supports the proposal of an anisotropic extended $s_{\pm}$-wave 
pairing in FeSe.}
\label{fig3}
\end{figure*}

USC with a finite $\Delta_{s_{\pm}}=\langle c_{ka\sigma}^{\dag}c_{-kb-\sigma}^{\dag}\rangle >0$ now occurs {\it in} the $C_{2v}$ structure.  This instability is studied 
exactly as before,~\cite{laad} by solving $H=H_{n}+H_{res}^{(MF)}$, but where 
$H_{n}$ is now the full five-band Hubbard model {\it including} ON+BOO 
order.~\cite{laad1} In particular, the ``normal'' state Fermi surface is 
now already reconstructed above $T_{c}$, and the $xz$-pocket is absent, 
in accord with very recent ARPES data.~\cite{rhodes} In Fig.~\ref{fig2}, 
we show the local spectral function (DOS), now interpreted as usual as the 
Bogoliubov quasiparticle spectrum in the SC state.  A number of features, 
germane to USC in FeSe, immediately stand out. First, as observed generically 
across strongly correlated SC,~\cite{sachdev} an anomalous ``normal'' state 
without LFL quasiparticles [finite Im$\Sigma_{a}(\omega=E_{F})$ with a 
quasilinear energy dependence] gives way to a much more coherent low-energy 
response V-shape at low energies in the gap region, strongly suggesting a 
nodal gap structure. $(ii)$ Spectral-weight redistribution across the SC 
transition on an energy scale much larger than the gap value (see below), 
testifying to the strongly correlated nature of the system.  Remarkably, 
opening up of the SC gap also results in orbital-selective sharpening of 
the low-energy coherent feature just below $E_{F}$ {\it and} to the 
(self-consistent) development of a pronounced peak-dip-hump structure 
that should be visible in spectral probes.  $(iii)$ That SC occurs in 
a ``normal'' state with broken $C_{4v}$ symmetry is manifest in the DOS 
asymmetry in the normal and USC states.  While $(i)$ and $(iii)$ are 
consistent with STM data,~\cite{song,loidl} $(ii)$ requires PES experiments 
to be analyzed over a larger energy window than hitherto done: that this 
will show up sizable spectral weight transfer and the orbital-selective 
peak-dip-hump features across $T_{c}$ is a specific prediction of our work.

The nodes in the gap function, implied by the V-shaped DOS in the USC state 
are rationalized by observing that the SC pair function with anisotropic
$s_{\pm}$-symmetry, $\Delta_{s_{\pm}}(k)=\Delta_{1}(c_{x}+\eta c_{y})+\Delta_{2}c_{x}c_{y}$ with $c_{x,y}=$cos$k_{x}$,cos$k_{y}$ and $\alpha=\Delta_{2}/\Delta_{1}\simeq O(0.7-1.0)$ will intersect the Fermi pockets in FeSe at the $M$ 
points.~\cite{chainani} Quite remarkably, the momentum-averaged spectral 
function (DOS) in the USC state turns out to be in good accord with 
PES data.~\cite{SC-FeSe-pes} as shown in Fig.~\ref{fig3}. Our computed 
results are also in fair accord with STM data,~\cite{song} clearly revealing 
$(i)$ the $C_{2v}$ lattice symmetry in which SC appears as an asymmetry in 
$\rho_{xz,yz}(\omega)$, $(ii)$ the overall spectral lineshape, as well as 
$(iii)$ the clear V-shaped gap in the DOS.  

Remarkably, our proposal also accounts for NMR~\cite{cava} data for FeSe 
as follows.  We have studied the $T$-dependent evolution of local dynamical 
spin fluctuations in the whole $T$-range by computing the nuclear magnetic 
resonance (NMR) spin-relaxation rate $1/T_{1}T$ as a function of $T$ in FeSe 
by repeating earlier calculations for the 1111-FeAs systems.~\cite{our-nmr} 
$(T_{1}T)^{-1}$ measures the weighted average of the various ${\bf q}$ modes 
of low-frequency spin fluctuations. Satisfyingly, as seen in Fig.~\ref{fig4}, 
correct $T$-dependence of $(T_{1}T)^{-1}$ over the whole $T$-range is well 
captured by DMFT. This clearly marks the strong enhancement of 
large-${\bf q}$ spin fluctuation modes as $T$ is reduced toward $T_{c}$. 
In the normal state, the source of the increasing spin correlations lies in 
the partially unquenched and strongly quantum-mechanically fluctuating 
local moment in DMFT: this is also what causes bad-metallic normal state 
conduction~\cite{craco-fese,biermann} in FeSe$_{1-x}$Te$_{x}$.  A normal state
pseudogap is clearly evidenced by the fact that $1/T_{1}T$ {\it decreases} 
below $T^{*}\simeq 25$~K($\simeq 2T_{c}$) in excellent accord with very 
recent estimates.~\cite{grinenko} Below $T_{c}$, $T_{1}^{-1}$ suddenly 
decreases due to the suppression of low-energy spin fluctuations by SC 
gap opening.  Absence of the Hebel-Slichter (HS) coherence peak is a 
consequence of a strongly incoherent normal state self-energy: the latter 
overdamps the coherence peak. Finally, $T_{1}^{-1} \simeq T^{n}$ with 
$3<n<5$ below $T_{c}$, in accord with in-plane SC-gap nodes: this will also 
result in vanishing of the HS peak.  Furthermore, using 
$\Lambda_{A}/T\simeq 1/T_{1}T$,~\cite{grinenko} we also estimate the muon 
depolarization rate in the $\mu$ SR study: the inset of Fig.~\ref{fig4} 
shows $1/T_{1}T$ vs $T$ on a  log-log plot. Clearly, both $1/T_{1}T$ and 
hence $\Lambda_{A}/T$ for $V_{ab}^2=0.021$ exhibit approximate power-law 
behavior with $T$, $i.e.$, they vary as $T^{-n}$ with $0<n<1$, in nice 
qualitative accord with data, though the exponent $n$ in the $T^{-n}$ 
dependence is same for both (probably, a reflection of our use of a 
$T$-independent $A(q)$~\cite{grinenko} in the calculation). Below 
$T^{*}=25$~K, the critical scaling behavior is cut-off by the low-energy 
pseudogap, again in full accord with data.  In our theory, this pseudogap 
(see Fig.~\ref{fig3}) emerges from enhancement of orbital-selective Mottness 
accompanying onset of ON+BOO order as above, and not from preformed 
Cooper pairs.    

\begin{figure}[!b]
\vskip50pt
\end{figure}
\begin{figure*}[!t]
\includegraphics[width=5.9in]{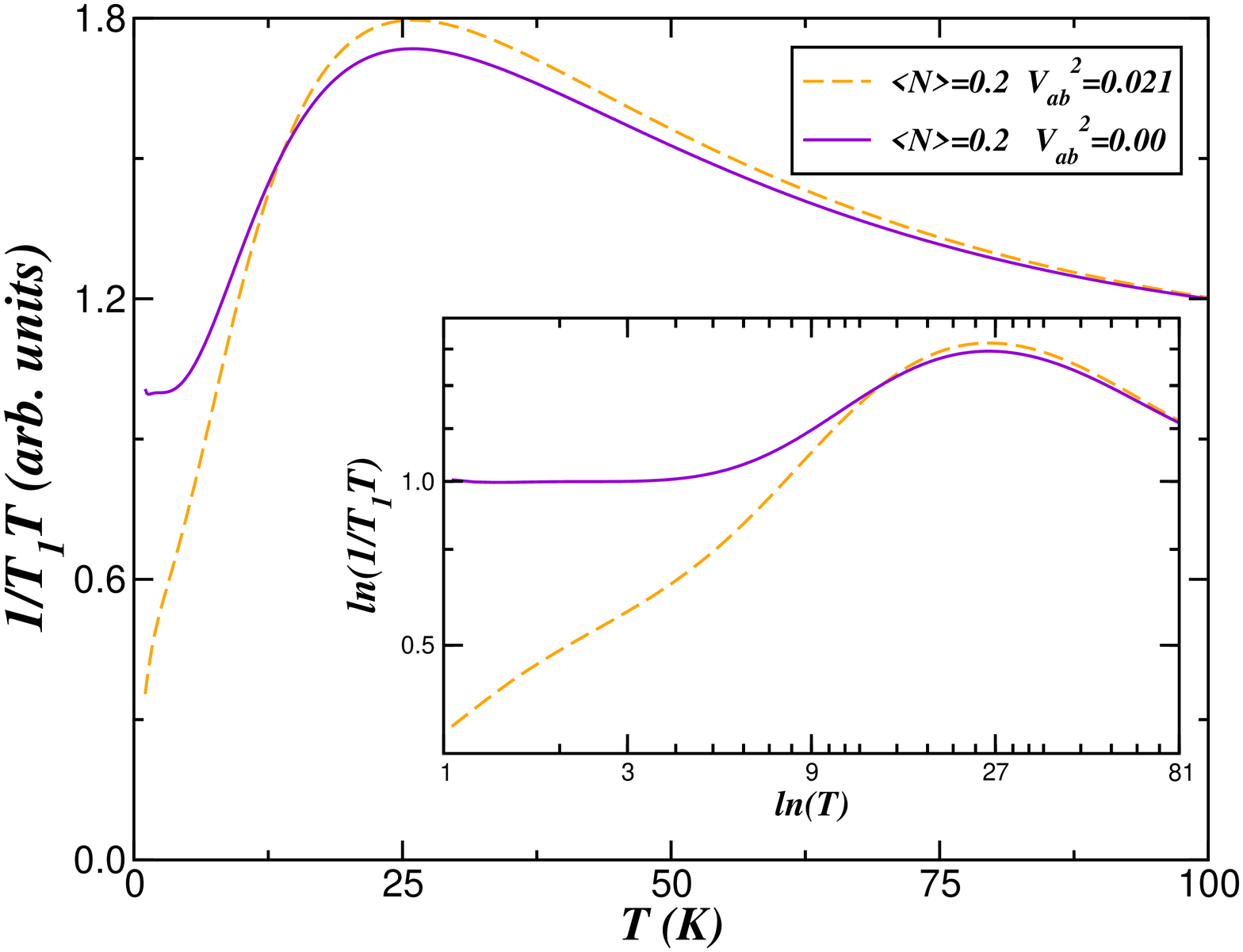}
\caption{
The $T$-dependent NMR spin-lattice relaxation rate for the normal and 
our proposed anisotropic extended $s_{\pm}$-wave SC in FeSe.  Direct 
comparison with data~\cite{cava} shows impressive accord over the whole 
$T$-range. Above $T_{c}$, $(T_{1}T)^{-1}$ shows a $T$-dependent increase, 
showing progressive build-up of short-range (large-${\bf q}$) AF spin 
correlations.  Absence of Hebel-Slichter peak above $T_{c}$ and power-law 
fall-off for $T<<T_{c}$ are signatures of strong normal state scattering
 and nodal SC gap (see text for explanations).}
\label{fig4}
\end{figure*}

These observations show that the ``pair glue'', arising from $H_{res}$, is 
essentially composed of entangled inter-site spin- and orbital (not 
spin-orbit) dynamical modes in the residual interactions ($H_{res}$), 
even as FeSe does not show AF order.  Since SC pairing is predominantly 
short-ranged (near and next-nearest neighbors in $H_{res}$), the pair 
coherence length will be small and the upper critical field, $H_{c2}$, 
will be large ($H_{c2}=47$~T under pressure,~\cite{awana}) as seen.  The 
``normal'' state  clearly has critical spin fluctuations, as evidenced 
by $1/T_{1}T\simeq T^{-n}$ for $40<T<80$~K in Fig.~\ref{fig4}.  We posit 
that such critical magnetic fluctuations can naturally arise from a lattice 
orthogonality catastrophe accompanying orbital-selective Mottness arising 
from the Lifshitz transition discussed before.  This is detailed in 
Supplementary Information (SI): interestingly, we find that the power-law 
$T$-dependencies  related to soft dynamical {\it spin}-quadrupolar (SQ) 
fluctuations (see SI), $i.e$, to a Fourier transform of 
$<S_{i}^{x}Q_{i}^{x}(\tau);S_{i}^{x}Q_{i}^{x}(0)>$, with $S_{i}^{x}Q_{i}^{x}=(c_{i,a,\sigma}^{\dag}c_{i,b,-\sigma}+h.c)$,
with $a=xz, b=yz$ in the $xz,yz$ orbital sector.  Given a finite FQO as 
derived before from $H_{res}$, a direct symmetry adapted coupling between 
FQO and SQ-fluctuations naturally arises: observation of critical scaling 
in $1/T_{1}T$ is now directly tied to the latter. Our findings are thus also 
consistent with a metallic state with FQO and SQ correlations, and not too 
far from AF order, which actually occurs under pressure.~\cite{grinenko} A 
detailed elucidation of this last point requires a study of FeSe under 
pressure, and is out of scope of this work.

\section{Discussion}
          
Given the form of $H_{res}$, the main competitor to SC is an orbital-EN-plus 
BOO or FQO state that drives a T-O structural transition.  Support for this 
comes from a $T-x$ phase diagram~\cite{loidl} and ARPES:~\cite{ding} in the 
former, SC $T_{c}$ peaks precisely at a $x_{c}$ where the T-O transition is 
suppressed to $T=0$.  This also has other consequences, outlined below. 
First, replacing Se by Te causes electronic structure changes owing to 
reduced chalcogen height ($h$) in the unit cell.  This parameter has 
been identified as relevant to the fermiology of FeAs systems,~\cite{belen}
and the sensitivity of LDA FS to $h$ is well known.  In FeSe geometry, 
decreasing $h$ increases the diagonal Fe-Te-Fe hopping relative to the 
near neighbor Fe-Fe hopping.  Translated into dependence of $J_{1},J_{2}$ 
on $h$ in our strong coupling view, this means an effective change in the 
effective residual two-particle interaction, pushing the ratio 
$J_{2}/J_{1}$ to values high enough that the SC gap function no longer 
intersects the renormalized FS in the $C_{2v}$ structure:~\cite{chainani} 
This will drive nodeless, anisotropic $s_{\pm}$, accounting for the transition 
from nodal to nodeless $s_{\pm}$ SC as seen.~\cite{song,chainani} Also, the 
suppression of SC under strain and its enhancement at high pressure finds 
a natural explication: strain couples linearly to the (orbital) nematic 
order parameter, and stabilizes nematicity. Since $\langle N\rangle$ 
competes with $\Delta_{s_{\pm}}$, this weakens $s_{\pm}$ SC.  And, as is 
ubiquitous in MO systems, high pressure weakens nematic order by reducing 
the $d_{xz}-d_{yz}$ splitting, enhancing SC.~\cite{miyoshi} Finally, minute 
impurity content (e.g, excess Fe in Fe$_{1+\delta}$Se with $0<\delta <<1$) 
will rapidly suppress nodal $s_{\pm}$ SC (in the parent FeSe) on rather 
general grounds. Thus, our proposal provides a natural qualitative 
explication for a wide range of unusual behaviors exhibited in the SC 
state in Fe$_{1+\delta}$Se$_{1-x}$Te$_{x}$. 

That USC arises in an orbital nematic state with $C_{2v}$ structure is also 
consistent with the fact that the observed vortex shape anisotropy strongly 
exceeds values expected on pure structural grounds.~\cite{song,subir1} 
Further, absence of AF order and interband nesting at 
${\bf Q}=\pm(\pi,0),\pm(0,\pi)$ in the correlated electronic structure 
presents problems for a magnetic origin for nematicity in FeSe.  But 
orbital EN-plus BOO does not require nesting, and thus emerges as the 
natural contender for the competing order in FeSe.  Given the loss of the $xy$-band states due to the OSMP in the ``normal'' state already above FQ and SC ordering scales also precludes use of LDA band structures and associated nesting
features to rationalize such features.

We emphasize that Cooper pairing naturally turns out to be orbital-selective 
in our work, since the ``normal'' state above $T_{c}$ already has enhanced 
orbital-selectivity due to finite $<Q_{i}^{z}>$, $i.e$, a state with FQO, 
also viewable as an orbital nematic.  This is manifest in our finding of 
orbital-selective anisotropy in the Bogoliubov DOS in the SC state above, 
which exhibits marked orbital selective changes with- and without a finite 
SC pairing, simply because such orbital differentiation is already driven by 
FQO {\it above} $T_{c}$. Orbital-selective pairing arises because, in such a 
state, our results imply that pairing in the more quasicoherent $yz$-band 
states is favored relative to that in the $xz,xy$ bands, which are pushed 
down by orbital-selective Mott ($d_{xy}$) and by FQO ($d_{xz}$): the former 
already obtains as a genuine correlation effect (orbital-selective Mottness) 
in the normal state without any broken symmetry, while the latter is a 
consequence of the downshift of the $xz$-FS pocket arising from $H_{res}$ 
in our work.  This is naturally consistent with the observation of 
orbital-selective spectral enhancement from $d_{yz}$ states at the expense 
of $d_{xy,xz}$ ones in recent quasiparticle interference studies,~\cite{sprau} 
where the spectra can be well-fitted by theoretically solving the linearized 
Eliashberg model {\it assuming} that only the $d_{yz}$ band contributes to SC 
pairing.  This is also consistent with recent ARPES data.~\cite{rhodes}  
Quantitative comparison between our work and these QPI and ARPES data requires 
more work, which we leave for future consideration.

Moreover, our proposal of FQO as a competing order (in the particle-hole 
channel) is consistent with electronic Raman scattering~\cite{blumberg} and 
NMR studies~\cite{grinenko} for the following reason.  Focussing on the 
$xz,yz$ orbital sector, one can define pseudospin-$Q^{\mu}=1/2,S^{\mu}Q^{\mu}=/2$ degrees of freedom in such a two-orbital system with $\mu=x,y,z$ as
$Q^{x}=(c_{xz,\sigma}^{\dag}c_{yz,\sigma}+h.c)/2, Q^{y}=(c_{xz,\sigma}^{\dag}c_{yz,\sigma}-h.c)/2i, Q^{z}=(n_{xz,\sigma}-n_{yz,\sigma})/2$, and $S^{x}Q^{x}=(c_{xz,\sigma}^{\dag}c_{yz,-\sigma}+h.c)/2$,
with $S^{y}Q^{y}, S^{z}Q^{z}$ defined analogously.  $Q^{z}$ is precisely the FQ 
order parameter, while the correlator of $Q^{x}$ describes FQ-fluctuations 
measured in Raman studies.  Simultaneously, the correlator of $S^{x}Q^{x}$ is 
now associated with {\it spin}-quadrupolar fluctuations on each Fe-site.  In 
the SI, we show that local spin-quadrupolar fluctuations exhibit power-law 
$T$-dependence due to a lattice orthogonality catastrophe, {\it a la} the 
strange metal, that is intimately tied down to an OSMP.  This is because 
the SQ-correlator is precisely the inter-orbital ``spin excitonic'' 
fluctuator, which {\it is} infra-red singular due to the venerated 
orthogonality catastrophe.  OSMP in the Fe-based SCs has been extensively 
discussed using various theoretical approaches, and occurrence of 
strange metallicity in 122-Fe arsenides is also well known.

Importantly, in our work, both, the FQO and USC naturally emerge as 
instabilities of the incoherent metal, originating from the same residual 
interaction ($H_{res}$). Hence, since both instabilities compete for the same 
Fermi pocket(s), it is clear that they are competing orders, consistent with 
STM and NMR~\cite{brink} data.  Theoretically, our work is distinct in spirit 
from dominantly itinerant views, where OSMP cannot appear by construction.  
Recently, Mukherjee {\it et al.}~\cite{peter} have studied EN and SC states 
in such an itinerant spin fluctuation view based on LDA Fermi surfaces, 
where both arise from a normal state with well-defined Landau quasiparticles.  
For ordered states, our work is thus complementary to theirs, in the sense 
that both approaches give the correct pair symmetry.  Very recent 
work~\cite{benfatto} propose that an orbital-dependent spin fluctuation 
mediated pairing, mediated by processes beyond RPA, is in good accord with 
ARPES and STM data.  While comparison of our approach with theirs is premature 
and requires more work, our proposal also naturally involves strongly entangled 
spin-orbital correlations, and the spin fluctuations are intrinsically 
orbital-selective.  This is explicit in the residual two-particle interaction, 
$H_{res}$.  The distinguishing features in our approach are $(i)$ strongly 
enhanced orbital-selectivity leads to an OSMP (loss of $xy$-band states) and leads to breakdown of
 Landau quasiparticle description already in the ``normal'' state, in favor 
of a metallic state with branch-cut responses in the correlation functions, 
a feature of the ``strange'' metal,~\cite{pwa} and $(ii)$ SC pairing in our 
case involves strong short-ranged, spin-orbital fluctuations, and will lead 
to a short coherence length SC pairing. High $H_{c2}$ values and the fact that 
FeSe is close to a BCS-BEC crossover~\cite{lohneysen} support such a view.

In our picture, however, an orbital-selective Mott behavior is the 
symmetry-unbroken ``normal'' state~\cite{craco-fese}, well-documented 
theoretically, is necessary: $(i)$ lack of hole pockets in the renormalized 
electronic structure (via the Lifshitz transition at $T^{*}$ in experiments) 
leads to breakdown of ``adiabatic continuity'' with LDA Fermi surfaces. 
It is then impermissible to use LDA Fermi surfaces to derive instabilities 
to ordered states, and $(ii)$ it is the vanishing LFL quasiparticle 
residue in the ``normal'' state ($z_{FL}^{xz}=0$) that leads to preferential 
relevance of the two-particle residual interaction, $H_{res}$, relative to 
the one-electron hybridization between $xz,yz$ states, $H_{hyb}^{(1)}$, in LDA.  
We have shown that {\it both}, FQO and USC orders follow from $H_{res}$ as 
competing orders, and discussed how this is qualitatively consistent with
indications from a wide variety of spectral probes. 

If the normal state were a weakly correlated LFL, $H_{hyb}^{(1)}$ would 
necessarily need to be preferentially relevant in FeSe in a situation where 
FS nesting is absent. This latter view would be applicable if a FL metal 
underwent conventional BCS-like instabilities to nematic and SC orders.  
Unambiguous observation of nLFL behavior (bad-metallicity and 
quasi-linear-in-$T$ resistivity) in the ``normal'' state in FeSe without 
any symmetry-breaking supports the approach taken here. Together with 
earlier DMFT work~\cite{craco-fese,kotliar} for Fe-based SC and the good 
accord with a range of key physical responses in both, normal and USC states, 
the present approach testifies to the relevance of strong correlation-driven 
orbital-selective-Mott state and its novel role in fomenting unconventional 
nematic and superconductive orders in FeSe.   

\mysection{Methods}
\noindent
To study the competing orders in  
FeSe, we employ the local-density-approximation plus 
dynamical-mean-field-theory (LDA+DMFT) which takes into 
consideration the most relevant local multi-orbital and associated Hund correlation effects germane to the strongly correlated metallic state in FeSe. LDA+DMFT~\cite{ourV2O3} 
implementation used here also correctly describes disorder, pressure and 
temperature effects in multi-band electronic systems. The LDA+DMFT scheme 
is an ideal starting point towards description of correlation-driven 
metal-to-insulator transitions, Fermi and non-Fermi liquid metallic 
states in quasi-local limits solely due to strong dynamical correlations 
in idealized many-particle models as well as in real multi-orbital 
systems.~\cite{kotrev} It also provides 
 mean-field descriptions of the competing orders in correlated systems at 
low temperatures, and of large modification of physical properties 
in response to external 
perturbations like pressure, chemical doping, magnetic and electric fields, 
etc. The one-electron LDA density-of-states were computed using the 
non-fully relativistic version of the PY-LMTO code.~\cite{craco-fese} 
To incorporate the effects of dynamical electronic correlations in FeSe, we use the multi-orbital 
iterated-perturbation-theory (MO-IPT) as an impurity solver of the 
many-particle problem in DMFT, as described in detail in 
Refs.~\onlinecite{ourv2o3}.

\mysection{Acknowledgements}
\noindent M.S.L. thanks the Helmholtz-Zentrum Berlin, Germany, and the DAE, 
Govt. of India, for support and hospitality during the time this work was 
done. L.C.'s work is presently supported by CNPq (Proc. No. 304035/2017-3). 

\mysection{Author contributions}
\noindent M.S.L conceived the problem.  L.C performed DMFT calculations.  M.S.L and L.C analyzed the LDA+DMFT results and wrote the manuscript with inputs from B.F. All authors discussed the results and reviewed the manuscript.

\mysection{Additional information}
\noindent {\bf Competing financial interests:} The authors declare no 
competing financial interests.\\
\noindent Correspondence and requests for materials should be addressed 
to M.S.L. (mslaad@imsc.res.in,tanguero7@hotmail.com).

\newpage 
%\vspace{2cm}

\section{Supplementary Information: Unusual Locally Critical Spin 
Fluctuations from Orbital Selective Mottness}

Here, we detail how correlations in an orbital-selective Mott phase (OSMP)
offer a natural insight into the intermediate-in-$T$ power-law behavior of 
spin fluctuations in FeSe.  To this end, we analyze the analytic structure of 
local, one- and two-particle correlators in the OSMP by appealing to the 
underlying impurity model of DMFT, where analytic progress is possible. We 
explicitly show how these are very distinct in the OSMP and non-OSMP phases 
of the model, the difference arising from the fact that a local orthogonality 
catastrophe in the OSMP destroys the Landau quasiparticles via a fundamental 
change in the analytic structure of the one-electron propagator from a pole- 
to an infra-red singular branch-cut structure at low energy.

In the text, we have shown how onset of an orbital nematic-plus bond-ordered
orbital (ON+BOO) state enhances the already substantial orbital-selective
correlations in FeSe.  Specifically, such correlations push the $xz$-band 
states {\it below} $E_{F}$ (Lifshitz transition), whence this metal now has 
co-existent ``metallic'' ($yz$-band) and Mott localized ($xz$-band) states at 
low energy.  We restrict ourselves to only the $xz,yz$ orbital sector to flesh 
out the essential physics.  The multi-orbital interaction terms in the 
two-band Hubbard model,

\be
H_{xz,yz}=U'\sum_{i,\sigma,\sigma'}n_{i,xz,\sigma}n_{i,yz,\sigma'} 
- J_{H}\sum_{i}{\bf S}_{i,xz}.{\bf S}_{i,yz}
\ee
compete with the inter-orbital one-electron hybridization, 
$H_{hyb}^{(1)}=t_{xz,yz}\sum_{i,j}(c_{i,xz,\sigma}^{\dag}c_{j,yz,\sigma}+h.c)$ 
(putting $a=xz, b=yz$ in $H_{hyb}$ in the main text).  In the OSMP phase, 
the interplay between $H_{xz,yz}$ and $H_{hyb}^{(1)}$ leads to emergence of 
anomalous behavior, which we now discuss.

Were the metallic state to be a Landau Fermi-liquid (LFL) metal, 
$H_{hyb}^{(1)}$ would be necessarily relevant in the renormalization group 
sense.  This would be the case only as long as no OSMP occurred in the DMFT 
solution, since a relevant inter-orbital one-electron hybridization requires 
absence of any Mott-like localization (since otherwise, no coherent 
one-electron mixing in the $xz-yz$ channel could result).  Once the OSMP 
sets in, this situation undergoes a basic change: In the OSMP, action of 
$H_{hyb}^{(1)}$ will necessarily mix a metallic band $yz$ electron with a 
Mott-localized $xz$-band electron. However, in this case, a coherent 
one-electron mixing would be suppressed to zero, since the lower Hubbard 
band states in the $xz$-band are now filled with one electron per Fe site, 
and so cannot recoil, simply because there are {\it no} lower-Hubbard band 
states into which they can do so.  In fact, the $xz-yz$ mixing will necessarily 
generate a doubly occupied state in the $xz$-sector, where a $yz$-band fermion, 
transferred to the $xz$-oprbital, will result in a {\it doubly occupied} state 
on the $xz$-orbital.  But this is a high-energy, upper-Hubbard band state, 
and {\it thus has no one-to-one correspondence with any one-electron state}. 
The upshot is that the corresponding mixing term thus undergoes a radical
change, and now reads $H_{hyb}'\simeq t_{xz,yz}'\sum_{i,\sigma}(n_{i,xz,-\sigma}c_{i,xz,\sigma}^{\dag}c_{j,yz,\sigma}+h.c)$, i.e, it describes hybridization of
metallic $yz$ states with upper-Hubbard band states in the $xz$ sector, which 
are {\it projected} $X$-fermions.  This correlation-induced blocking of the 
one-electron hybridization thus leads to irrelevance of$H_{hyb}^{(1)}$ and to 
Kondo destruction, and results in a non-LFL state.

The mechanism of breakdown of Landau FL metallicity is now manifest. The above 
argument leads us to consider the two-orbital Hubbard model in a regime where
$H_{hyb}^{(1)}$ is irrelevant, and so the physics is now controlled by strong 
scattering between $xz,yz$ states.  As stated above, the fact that 
$xz$-carriers cannot any more recoil during scattering leads to very 
interesting anomalies. [S1]  Specifically, within the local `
`impurity problem'' of the DMFT, both interaction terms in $H_{xz,yz}$ 
now correspond to strong scattering of $yz$ carriers by recoil-less (Mott 
localized) $xz$-states.  This corresponds precisely to the venerated 
X-ray edge problem, [S1,S2] where the $xz$-fermion propagator as well 
as the ``excitonic'' inter-band fluctuation propagator acquires anomalous 
dimensions, and the one- and two-particle spectra exhibit anomalous 
branch-cut continuua instead of the conventional pole structure of a 
LFL metal: explicitly,

\be
\rho_{xz}(\omega)=\int d\tau e^{i\omega\tau}<T_{\tau}c_{xz}(\tau)c_{xz}^{\dag}(0)> \simeq \theta(-\omega)|\omega|^{-(1-\eta)}
\ee
with $\eta=\frac{1}{\pi}$tan$^{-1}(U'/W)$, where $W$ is an effective one
electron band-width.  Thus, $0<\eta <1$. In Fig.~\ref{fig1}, we indeed 
find that in the ON+BOO phase (with $<N>=0.2$), the local spectral function 
of $xz$-states shows a very deep pseudogap (a finite-temperature effect, 
we would have $\rho_{xz}(\omega=E_{F})=0$ at $T=0$), accompanied by an 
anomalously broadened lineshape of half-width $O(1)$~eV, while the 
corresponding $yz$-spectral function exhibits clear power-law fall-off 
of the above type with no discernible Landau quasiparticle pole-like peak 
at $E_{F}(=0)$, in full qualitative accord with the above arguments.

More crucially, in such a ``strange'' metal, the corresponding local spin 
fluctuation spectrum measured in NMR relaxation rate via 
$1/T_{1}T \simeq \lim_{\omega\rightarrow 0}\frac{\chi_{loc}''(\omega)}{\omega}$ 
with $\chi_{loc}''(\omega)$ being the imaginary part of the local dynamical 
spin-flip correlator, is also expected to show infra-red singular behavior. 
This is because the ``excitonic'' spin-flip correlation function involving 
``spin excitons'' created from the $xz,yz$ fermions, also has an anomalous 
branch-cut form due to the X-ray-edge mapping:

\bn
\nonumber
\chi_{loc}''(\omega) &=& \int d\tau e^{i\omega\tau}<T_{\tau}c_{i,xz,\sigma}^{\dag}c_{i,yz,-\sigma}(\tau);c_{i,yz,-\sigma}^{\dag}c_{i,xz,\sigma}(0)> \\ 
&\simeq& |\omega|^{-(2\eta-\eta^{2})}
\en

Remarkably, this ``interband excitonic'' correlation function is precisely 
that of a {\it spin quadrupole}, defined as 
$Q_{i,x}^{\sigma,-\sigma}=(c_{i,a,\sigma}^{\dag}c_{i,b,-\sigma}+h.c)/2$.
In the $xz,yz$-degenerate orbital sector relevant to undistorted structure of 
FeSe, it is natural to define orbital pseudospins, $Q_{i,z}=(n_{i,a}-n_{i,b})/2, Q_{i,x}=(c_{i,a}^{\dag}c_{i,b}+h.c)/2, Q_{i,y}=(c_{i,a}^{\dag}c_{i,b}-h.c)/2i$, satisfying 
the usual pseudospin-$1/2$ SU(2) algebra: the first defines the FOO or 
FQO~\cite{blumberg} order parameter, while the second and third describe 
quadrupolar charge and current fluctuations in the FQO state.  In fact, in a 
two-orbital model, spin-orbital entanglement also implies 
{\it spin quadrupoles}, given by $Q_{i,x}^{\sigma,-\sigma}$ above, and fluctuations 
in the former will inevitably cause SQ fluctuations.  In the OSMP which is 
enhanced in the FOO~(FQO) state as shown in the main text, we have thus 
shown that a lattice orthogonality catastrophe results in singular 
SQ-fluctuations.  Thus, the power-law behavior in $1/T_{1}T$ and $\mu$SR 
relaxation rate arises from such singular correlations in the OSMP.  Our 
finding of a FQO state is also qualitatively consistent with recent Raman 
scattering work.~\cite{blumberg}

At finite $T$, this should yield $\omega/T$ scaling.  It would be interesting 
to test whether inelastic neutron scattering lineshapes show approximate 
$\omega/T$-scaling for FeSe {\it in the range where $1/T_{1}T$ exhibits a 
critical power-law behavior}: if these spin fluctuations would be almost 
momentum independent, this would lend further positive support to our idea. 
Using the above form for $\chi_{loc}''(\omega)$ to compute $1/T_{1}T$ immediately 
yields a $T^{-n}$ power-law behavior with $0<n<1$, as seen.  It is not, 
however, possible to quantitatively obtain the value of $n$ from this 
argument, because lattice self-consistency effect will affect the precise 
values of the above exponents (which correspond to the underlying ``impurity'' 
model).  Nonetheless, this argument directly rationalizes the power-law 
behavior of $1/T_{1}T$ above the pseudogap scale, and links it to an 
Anderson-Nozieres orthogonality catastrophe that arises naturally in 
an OSMP view.

On the other hand, in the non-OSMP state, which occurs in the $2$-band Hubbard 
model whenever the one-electron hybridization is relevant, the $xz$-fermion 
can {\it recoil} during the scattering processes involving $xz,yz$ carriers, 
now simply because absence of Mott localization enables such low-energy recoil 
to occur.  The impurity problem of DMFT for the $2$-band Hubbard model is now 
the above impurity model with a relevant $H_{hyb}^{(1)}=t_{ab}\sum_{<i,j>,\sigma}(c_{i,xz,\sigma}^{\dag}c_{j,yz,\sigma}+h.c)$. As is well known, a two-stage spin-orbital 
screening now quenches the local spin-orbital moments via orbital and spin- 
Kondo effect, leading to a correlated spin-orbital Kondo resonance in both,
$G_{xz}(\omega)$ and $G_{yz}(\omega)$ at low energy, reinstating Landau FL 
quasiparticles and a FL metallic state at low $T$.  Thus, within local (DMFT) 
approaches, observation of non-Landau quasiparticle behavior in the normal 
state requires an OSMP to occur.  Such an OSMP, enhanced by nematicity 
(or FQO in the main text) in FeSe, is also consistent with very recent QPI 
and ARPES results in the USC state in FeSe, as emphasized in the main text.

Finally, in the OSMP state, onset of orbital-nematic plus bond-order (ON+BOO) 
in FeSe further enhances orbital-selective correlations, opening up a 
low-energy pseudogap in $\rho_{yz}(\omega)$ as shown in the main text.  This 
suppresses the infra-red singular behavior described above, qualitatively 
accounting for the breakdown of QC scaling below $T^{*}\simeq 25$~K in our 
LDA+DMFT calculations, again in good accord with data.~\cite{grinenko}

[S1] Anderson, P. W. The 'strange metal' is a projected Fermi liquid with 
edge singularities. {\it Nature Phys.} {\bf 2}, 626 (2006).

[S2] Nozieres, P. $\&$ de Dominicis, C. T. Singularities in the x-ray 
absorption and emission of metals. III. One-body theory exact solution.
{\it Phys. Rev.} {\bf 178}, 1097 (1969).

\end{document}